# Investigation of ephaptic interactions in peripheral nerve of sheep using 6 kHz subthreshold currents

James Hope[1,2], Narrendar Ravi Chandra[1], Frederique Vanholsbeeck[2,3], Andrew McDaid[1]

[1]The Department of Mechanical Engineering, The University of Auckland, Auckland 1010, New Zealand
[2]The Dodd Walls Centre for Photonic and Quantum Technologies, Auckland 1010, New Zealand
[3]The Department of Physics, The University of Auckland, Auckland 1010, New Zealand

**Abstract**

The objective of this work was to determine whether application of subthreshold currents to the peripheral nerve increases the excitability of the underlying nerve fibres, and how this increased excitability would alter neural activity as it propagates through the subthreshold currents. Experiments were performed on two Romney cross-breed sheep *in vivo*, by applying subthreshold currents either at the stimulus site or between the stimulus and recording sites. Neural recordings were obtained from nerve cuff implanted on the peroneal or sciatic nerve branches, while stimulus was applied to either the peroneal nerve or pins placed through the lower hindshank. Results showed that subthreshold currents applied to the same site as stimulus increased excitation of underlying nerve fibres ($p < 0.0001$). With stimulus and subthreshold currents applied to different sites on the peroneal nerve, the primary CAP in the sciatic displayed a temporal shift of -2.5 to -3 µs which agreed with statistically significant changes in the CAP waveform ($p<0.02$). These findings contribute to the understanding of mechanisms in myelinated fibres of subthreshold current neuromodulation therapies.

## 1 Introduction

Neuromodulation therapies provide an alternative treatment modality for several drug-resistant neurological conditions, including Parkinson's, epilepsy, and depression [1-6]. While neuromodulation therapies generally administer suprathreshold currents to modulate activity in target neural cells, a smaller number administer subthreshold currents to either suppress or promote neural activity through partial hyperpolarization or depolarization, respectively, of cell membranes. For example, chronic subthreshold cortical stimulation for epilepsy [7, 8]; a high frequency (10 kHz) variant of spinal cord stimulation for pain management [2]; the conditioning current which precedes the suprathreshold current in transcranial magnetic stimulation for motor cortex studies [9]; and, transcranial current stimulation for depression [6].

Researchers have demonstrated this partial polarization effect on the vestibular system by administering subthreshold current, with a band pass filtered random noise waveform, to the mastoid processes and observing subject sway responses which were highly coherent with the applied current polarity and magnitude [10-12]. This coherence was explained by stochastic resonance, wherein subthreshold components of periodic stimulus and random noise stimulus sum to become suprathreshold [13, 14]. In two further studies [15, 16], when researchers applied a subthreshold current transcutaneously to the tibial nerve through electrodes placed proximal to the ankle, again with a band pass filtered random noise waveform, participants reported an increase in sensitivity to vibration applied to the foot. The authors of these latter two studies did not identify the underlying mechanism causing increased sensitivity, but postulated in one [16] that the applied current might increase synchrony of the sensory receptor action potentials in the tibial nerve due to augmentation of ephaptic interactions between active nerve fibres.

In ephaptic interactions, spatiotemporal variations in the electric field around an active neural cell influences activity in nearby neural cells by altering their membrane potentials. These interactions can improve synchrony of activity within populations of neural cells or triggering of neural activity in subthreshold cells, and may be induced either by naturally occurring physiological effects [17-19] or by artificially increasing excitability using chemicals [20, 21] and subthreshold electrical current [18, 22, 23]. In myelinated nerve fibres, modelling studies predict ephaptic interactions alter the propagation velocities of action potentials in neighboring active fibres which improves synchrony [24-27], and an *in vivo* study on rat showed increased activity in response to an electrical stimulus when it was temporally coupled with a compound action potential (CAP) [28]. Increasing synchrony and localized triggering of new action potentials in peripheral nerves using a subthreshold current presents an exciting prospect because such a paradigm could improve the signal to noise ratio in peripheral nerve interfaces, aid physical rehabilitation after spinal cord injury and stroke, and provide insight into mechanisms of subthreshold current neuromodulation therapies.

In the current study, we investigated augmentation of neural activity in hind limb of sheep *in vivo* by applying subthreshold, 6 kHz, sinusoidal currents to the peroneal nerve between stimulus and recording sites. This paradigm differs significantly from the *in vivo* rat study in [28] which evaluated changes in excitability induced by surrounding neural activity, and is more similar to the stochastic resonance experiments in [15, 16], though in the current study subthreshold currents were administered via nerve cuffs instead of transcutaneous electrodes, and neural recordings were acquired. Two evoked stimulus sites were employed, one via a nerve cuff implanted on the peroneal nerve, and the second via pin electrodes placed distal to the hock. We hypothesized that (1) administering subthreshold currents would increase excitability of the underlying fibres, and (2) this increased excitability would increase synchrony in the neural activity and alter the propagation velocities.

## 2 Materials and Methods

*2.1 Experiment apparatus*

Nerve cuff electrode arrays were fabricated from stainless steel foil and silicon using the method described in [29]. The 28-channel nerve cuff contained a 2 rings of 14 electrodes, spaced 7 mm apart, with 0.46 x 3 mm active area on each electrode, Fig. 1a. Electrodes were coated with poly(3,4-ethylenedioxythiophene):p-toluene sulfonate (PEDOT-pTS) to reduce the electrode-tissue contact impedance. The 2-channel nerve cuffs contained 2 electrodes, each 9.5 x 1 mm, spaced 6 mm apart, Fig. 1b. The 28 channel and 2 channel electrode arrays were glued (Smooth-On Sil-Poxy®) into elastomer cuffs with dimensions (L x O.D. x I.D) of 20 x 8 x 4.5 mm and 6 x 6 x 3 mm, respectively, and with a slit along one side to allow implant on the nerve.

The 28-ch nerve cuff used for recording was connected to a headstage (INTAN C3314) via an adaptor (INTAN C3410), Fig. 1c. Neural recordings were acquired from each of the 14 electrodes on one electrode ring, with the electrodes of the other ring shorted together and used as the reference. Data were low pass filtered at 5 kHz for anti-aliasing, sampled at 20 kS/s, software notch filtered at 50 Hz to remove mains noise, then streamed via a USB interface board (INTAN C3100) to host PC and saved as .rhd files for processing later.

Stimulation was administered either through the pins placed along the cannon bone, or through electrodes in the most distally implanted nerve cuff on the peroneal nerve, and was generated by a bench-top pulse stimulator (AM-systems 2100) when digitally triggered (National Instruments CompactRIO® NI9403), Fig. 1c. The digital trigger was recorded on the USB interface board (INTAN C3100).

Subthreshold currents were generated using a PCB with parallel current-source circuits, developed by *The EIT Research Group at University College London,* and available for download from https://github.com/EIT-team. Currents were switched on and off using solid state relays (IXYS CPC1017N) controlled via digital lines (National Instruments CompactRIO® NI9403), Fig 1c.

*2.2 Tissue preparation and handling*

All animal procedures were approved by The University of Auckland Animal Ethics Advisory Committee. In total, two Romney cross breed sheep, female, and weighing 61 and 67 kg, were used in *in-vivo* experiments. Anesthesia was induced using by intravenous injection, and maintained using a mixture of isoflurane, oxygen and medical air administered via endotracheal tube. At the conclusion of experiments subjects were euthanized.

To implant nerve cuffs, once subjects were anesthetised, the left hind leg was extended and loosely fixed in position around the hock and fetlock joint. An incision was made down the posterior side of the thigh to expose underlying muscle, then the Semitendinosus and Biceps Femoris muscles were parsed apart, and the sciatic, tibial, and peroneal nerves were isolated by cutting away adipose tissue [30]. In subject 1, two 28-channel nerve cuffs were implanted adjacent to one another on the peroneal nerve, Fig. 1d. In subject 2, a 28-channel nerve cuff was implanted on the sciatic nerve, and four 2-channel nerve cuffs were implanted adjacent to one another on the peroneal nerve, Fig 1e. After implantation the muscle cavity was filled with physiological saline, preheated to 38 ºC, to cover the nerve cuffs and exposed nerve. A stainless steel pin was placed in this saline which connected via a 460 kΩ resistor to ground. In both subjects, two pins were placed 100 mm apart, subdermally, along the cannon bone to allow stimulation of distal sections of the peroneal nerve.

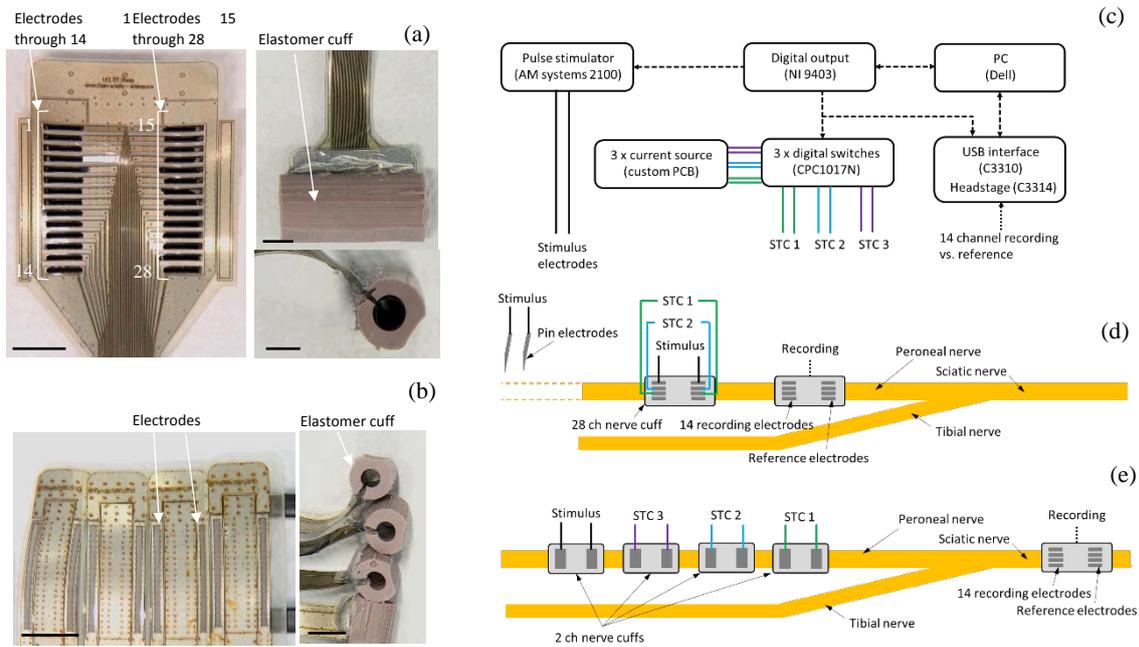

Figure 1: The 28 channel electrode array with two columns of 14 electrodes after PEDOT coating (left) and assembled into an elastomer cuff (right) (a), and four 2 channel electrode arrays before (left) and after (right) assembly into an elastomer cuff (b); black scale bars are 5 mm. A schematic of the experiment apparatus with solid lines for current controlled signals, dashed for digital lines, and dotted for analogue voltage recording (c), where STC = subthreshold current. Schematic of nerve cuff configuration in subject 1 (d) showing two 28-channel cuffs on the peroneal nerve, and in subject 2 (e) showing four 2 channel cuffs on the peroneal nerve and a 28-channel cuff on the sciatic nerve.

*2.3 Experiment protocol*

The subthreshold currents were biphasic, 6 kHz, sinusoidal waveform, and, in all but one protocol, +/- 70 µA amplitude. A 6 kHz sinusoid was selected to be sufficiently above the frequency components of neural activity to be filtered out during signal processing, and because transient impedance studies have indicated that 6 kHz results in more resistive current across the node of Ranvier and less capacitive current across the myelin sheath than in the neighboring frequencies [31, 32].

The amplitude of the stimulus pulses were selected to be 50 % higher than that which produced onset of twitching of the lower hindshank and phalanges, respectively, for stimulus administered via nerve cuff on the peroneal nerve and pins apposing the cannon bone. Twitching onset was characterized by manually triggering stimuli and increasing amplitude while visually monitoring the hind limb.

2.3.1 Hypothesis 1: Subthreshold current contribution to excitation of fibres

In subject 1, the contribution of subthreshold currents to excitation of fibres was tested by administering a pulse +/-0.3 mA, 50µs/phase, biphasic, square pulse stimulus every 500 ms to the distal 28-ch nerve cuff, Fig. 1d, through electrodes E11 and E23, Fig. 1a. In conjunction with this stimulus and on the same nerve cuff, no subthreshold current was applied for 5 seconds, then one subthreshold current was applied through E1 and E15 for 5 seconds, then a second subthreshold current – in phase with the first - was added through E4 and E18 for 5 seconds, Fig. 1a. This protocol was performed for a total of 600 seconds. The in-phase nature of the two currents produces a sinusoid with a +/- 140 µA amplitude within the cuff.

2.3.2 Hypothesis 2: Pin stimulus

In subject 1, augmentation of neural activity using subthreshold currents was investigated using stimulus applied through two pins inserted in the hindshank and apposing the cannon bone. Here, a +/-10 mA, 1 ms/phase, biphasic, square pulse was administered, while subthreshold currents were applied and recordings acquired from the peroneal nerve using the protocol described above in section 2.3.1. This protocol was performed for a total of 600 seconds with both stimulation pins inserted on the posterior side of the cannon bone, then repeated for a total of 600 seconds with the anodic pin inserted on the anterior side of the cannon bone. Finally, the protocol was repeated for a total of 100 seconds with the stimulus duration halved to 0.5 ms/phase to distinguish neural activity and stimulus artefacts in the neural recordings.

2.3.3 Hypothesis 2: Nerve cuff stimulus

In subject 2, augmentation of ephaptic interactions in the presence of three subthreshold currents was investigated using stimulus applied through a nerve cuff. Here, a +/-0.3 mA, 100 µs/phase, biphasic, square pulse was administered every 500 ms to the distal most 2-channel nerve cuff on the peroneal nerve, while no subthreshold current was applied for 5 seconds, then subthreshold currents were added one at a time, for 5 seconds, to each of the three adjacent 2-channel nerve cuffs, Fig. 1e. This protocol was repeated for a total of 600 seconds. The adjacent nature of the 2 channel nerve cuffs means the sinusoid amplitude did not exceed +/- 70 µA at any point along the peroneal nerve. Finally, to verify the influence of subthreshold current amplitude on the CAPs, the above protocol was repeated for a total of 600 seconds with the subthreshold current amplitude reduced to +/- 30 µA.

*2.4 Data processing*

Recorded data were processed in MATLAB (R2018b Mathworks). Data were parsed into 500 ms duration segments beginning with the digital trigger to the pulse stimulator, grouped into sets based on the number of applied subthreshold currents, and then notch filtered at 6 kHz to remove artefacts from the subthreshold currents. Signals with significant drift or corruption by artefacts, defined as having a mean value outside the bounds of -100 to 100 µV in the temporal window 20 to 30 ms, were removed.

Two metrics were used to evaluate changes in neural activity caused by the subthreshold currents: CAP amplitude, and CAP temporal shift. (i) CAP amplitude, defined as the difference between the maxima of the peak and the minima of the adjoining trough, was calculated for each of the data segments then compared for each subthreshold current condition using unpaired t-tests. (ii) CAP temporal shift was analysed by up-sampling the mean recorded waveforms for each current condition by a factor of 100 using cubic spline interpolation to increase temporal resolution, then comparing the upsampled waveforms using crosscorrelation. In addition, CAP amplitude and CAP temporal shift were evaluated by calculating the difference in mean recorded waveforms, and identifying points outside the +/-3 sigma noise threshold.

## 3 Results

*3.1 Surgery and nerve cuff implantation*

The femoral artery crossed over the sciatic, tibial and peroneal nerve branches, Fig. 2a, which hampered isolation of nerve branches from surrounding adipose tissue, and implantation of the nerve cuffs. The peroneal nerve was larger in subject 1, at 4 – 5 mm diameter, than in subject 2, at 3 mm diameter, which was one reason that compelled the use of different nerve cuff configurations between the two subjects Fig. 2b-c.

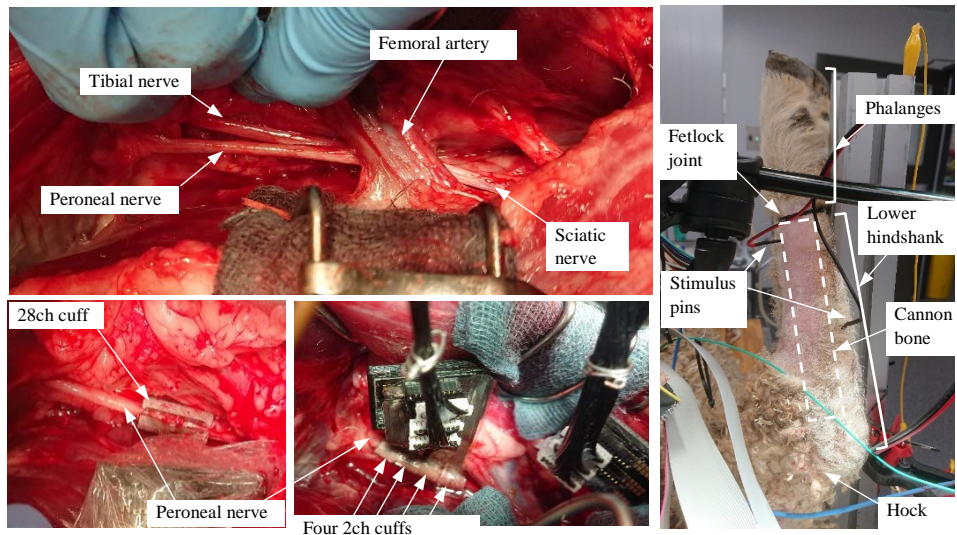

Figure 2: The sciatic, tibial, and peroneal nerve branches and femoral artery within the muscle cavity prior to nerve cuff implantation (a). One 28 channel nerve cuff implanted on the peroneal nerve in subject 1 (b), and four 2 channel nerve cuff implanted on the peroneal nerve in subject 2 (c). The hind limb fixed in place around the fetlock joint and hock, with stimulus pins visible in the lower hindshank (d).

*3.2 Hypothesis 1: Subthreshold current contribution to excitation of fibres*

Twitching of the lower hindshank was observed in response to the stimulus pulse, Fig. 2d. With two subthreshold currents applied, sustained extension of the lower hindshank was observed, indicating prolonged application of a +/- 140 µA amplitude, 6 kHz sinusoid is sufficient to activate motor fibres in the peroneal nerve which innervate the gastrocnemius muscle.

In all three current conditions – none, one, and two subthreshold currents – a CAP was observed between 0.3 and 0.8 ms, and with a peak at 0.4 to 0.45 ms, after commencement of the stimulus pulse. In electrode 8, which exhibited the largest CAP amplitudes, amplitudes of (mean +/- standard deviation): 1.58 +/- 0.33, 1.99 +/- 0.49, and 2.22 +/- 0.56 mV, respectively, were observed for none, one, and two subthreshold current conditions, Fig. 3a-c. Comparing these values with one another using unpaired t-tests (N = 300) produced two-tailed p values of < 0.0001, which is considered extremely statistically significant. Cross correlation produced lag values of 0 for both subthreshold current conditions. The difference in mean recorded waveforms were not analysed for these data.

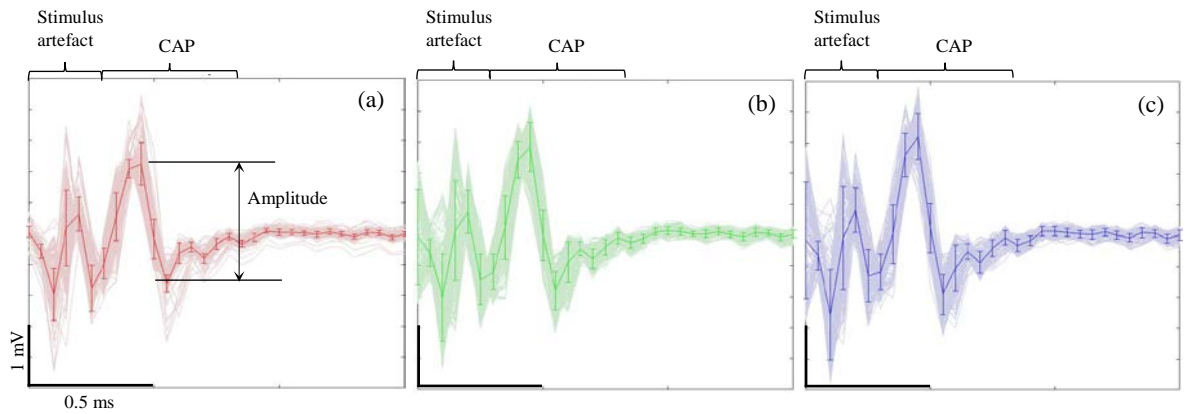

Figure 3: CAPs recorded from the peroneal nerve in response to stimulus applied, 30 mm distally, to the same nerve. Individual voltage recordings (shaded lines) overlaid with the mean and standard deviation (bold lines with error bars) with none (a), one (b), and two (b) subthreshold currents applied show that the subthreshold currents contribute to excitation of nerve fibres when applied to the same region of nerve as the stimulus pulse.

*3.3 Hypothesis 2: Pin stimulus*

Minor twitching of the phalanges were observed in response to the stimulus pulse. As was the case earlier, with two subthreshold currents applied sustained extension of the lower hindshank was observed.

The 1 ms/phase, biphasic, square pulse produced a stimulus artefact with 2.5 ms artefact from the pulse, followed by 1.5 ms of ringing, Fig. 4a-c. The shorter duration, 0.5 ms/phase, pulse produced a stimulus artefact with 1.5 ms from the pulse followed by 1.5 ms of ringing, Fig. 4d. With both stimuli, CAPs were observed from 4.5 ms onwards, indicating the artefact and CAPs could be distinguished form one another.

In both pin configurations and for all three subthreshold current conditions – none, one, and two subthreshold currents – neural activity was observed between 4.5 and 12 ms after commencement of the stimulus pulse, and contained multiple peaks and troughs with combined amplitudes in the range of 5 to 20 µV, Fig. 4b-c. CAP amplitudes were not analysed in the individual segments because they were obscured by noise. In the mean waveforms, where CAPs were visible, no changes in CAP amplitudes were identified above the 3σ noise threshold because of the large noise, of σ = 0.8 to 2.2 µV, relative to the CAP amplitudes, Fig. 4e-g. Cross correlation produced lag values of 0 for both subthreshold current conditions. These results indicate no changes in the CAP amplitude or temporal shift were identified outside the noise threshold.

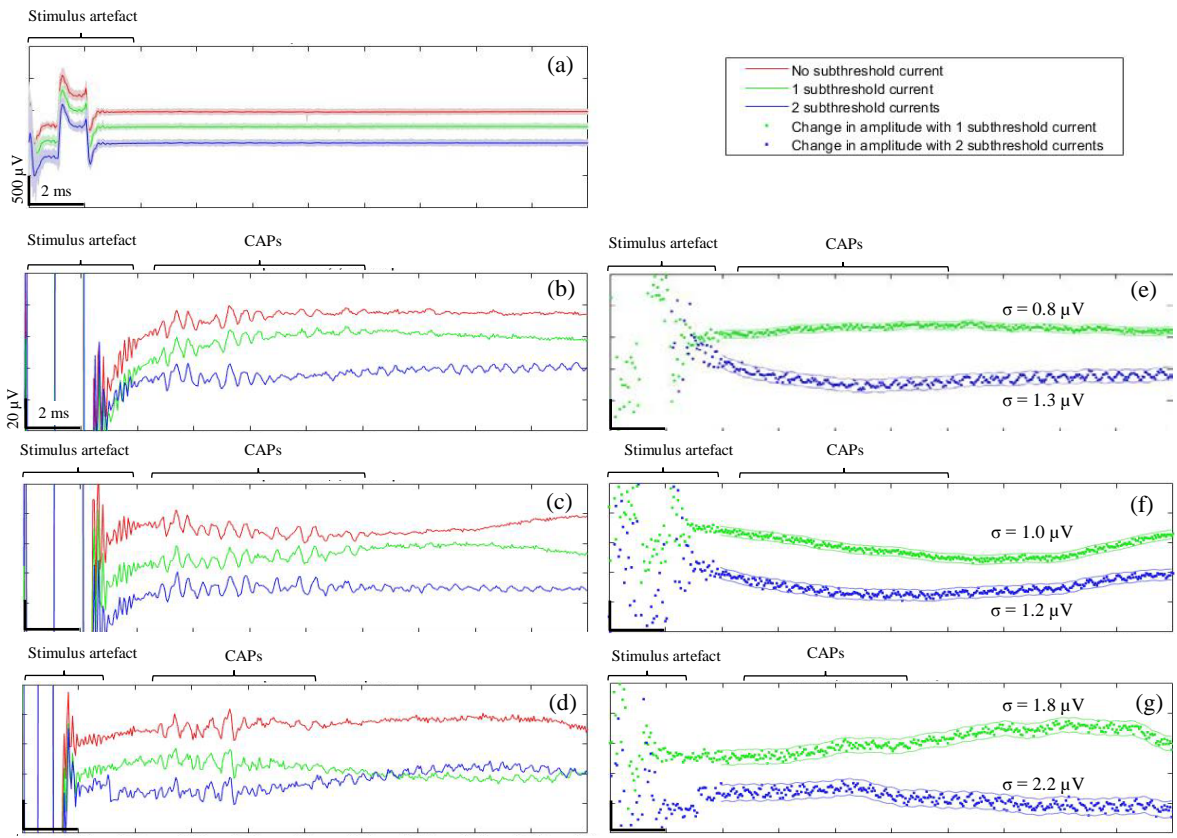

Figure 4: CAPs recorded from the peroneal nerve in response to stimulus applied approximately 300 mm distally through two pins apposing the cannon bone; with pins either on the posterior side of the cannon bone (a – b), or the posterior and anterior side of the cannon bone (c – d). The individual (shaded lines) and mean recordings (bold lines) show a large stimulus artefact between 0 and 4 ms (a). A close up of the mean recordings for none, one, and two subthreshold currents in red, green, and blue, respectively, show multiple CAPs with 5 to 20 µV amplitudes between 4.5 and 12 ms with a 1 ms/phase stimulus pulse (b-c), or 4.5 to 10.5 ms with a 0.5 ms/phase stimulus pulse (d). No changes in CAP amplitudes can be seen outside the 3σ noise threshold because of the large noise, of σ = 0.8 to 2.2 µV (e-g).

*3.4 Hypothesis 2: Nerve cuff stimulus*

In all four current conditions – none, one, two, and three subthreshold currents – a stimulus artefact was visible between 0 and 0.35 ms, followed by a large CAP with a mean amplitude of 2.13 to 2.14 mV and a peak at 0.7 ms after commencement of the stimulus pulse, Fig. 5a. Two smaller, secondary CAPs followed, with mean amplitudes of 13 +/- 1 and 35 +/- 1 µV and peaks at 2.85 +/- 0.05 and 5 +/- 0.05 ms, respectively, Fig. 5b. Lastly, a long duration CAP between 10 and 200 ms was visible with an amplitude between 59 to 61.5 µV and a peak at 50.9 to 53.3 ms, Fig. 5c. Multiple spikes were visible within the 10 to 400 ms temporal range, each with amplitudes of 10 to 400 µV and durations of 0.1 to 1 ms. The number and temporal location of the spikes varied between individual data segments, and were suspected to be caused by movement artefacts and evoked action potentials within the muscle fibres.

The primary CAP amplitude was (mean +/- standard deviation): 2.141 +/- 0.086, 2.136 +/- 0.088, 2.130 +/- 0.096, and 2.140 +/- 0.089 mV, respectively, for each of the four current conditions, Fig. 5a. Comparison of these values using unpaired t-tests (N=300) produced two-tailed p values between 0.13 and 0.83, which are considered to be not statistically significant. For the one, two, and three subthreshold currents, lag values of -5, -5 and -6 were calculated from cross correlation of the upsampled waveforms within the temporal window of 0.5 to 2 ms, corresponding to a temporal shifts of -2.5, -2.5 and -3 µs. These temporal shifts were visible in the difference in mean recorded waveforms, where the one, two, and three subthreshold current conditions all showed a positive difference at 0.65 ms followed by a negative difference at 0.75 ms, each with amplitudes of 60 to 80 µV, Fig. 5d. Using the unpaired t-tests (N=300), both the positive and negative peak difference amplitudes produced statistically meaningful p-values of 0.02 or less, despite large uncertainty at these points of 280 to 350 µV.

The amplitudes of the secondary CAPs and long duration CAPs were not analysed in the individual segments because they were obscured by noise. In the mean waveforms, where CAPs were visible, no changes in CAP amplitudes were identified above the 3σ noise threshold, Fig. 5e-f. Cross correlation of the mean waveforms produced lag values of 0 for all current conditions.

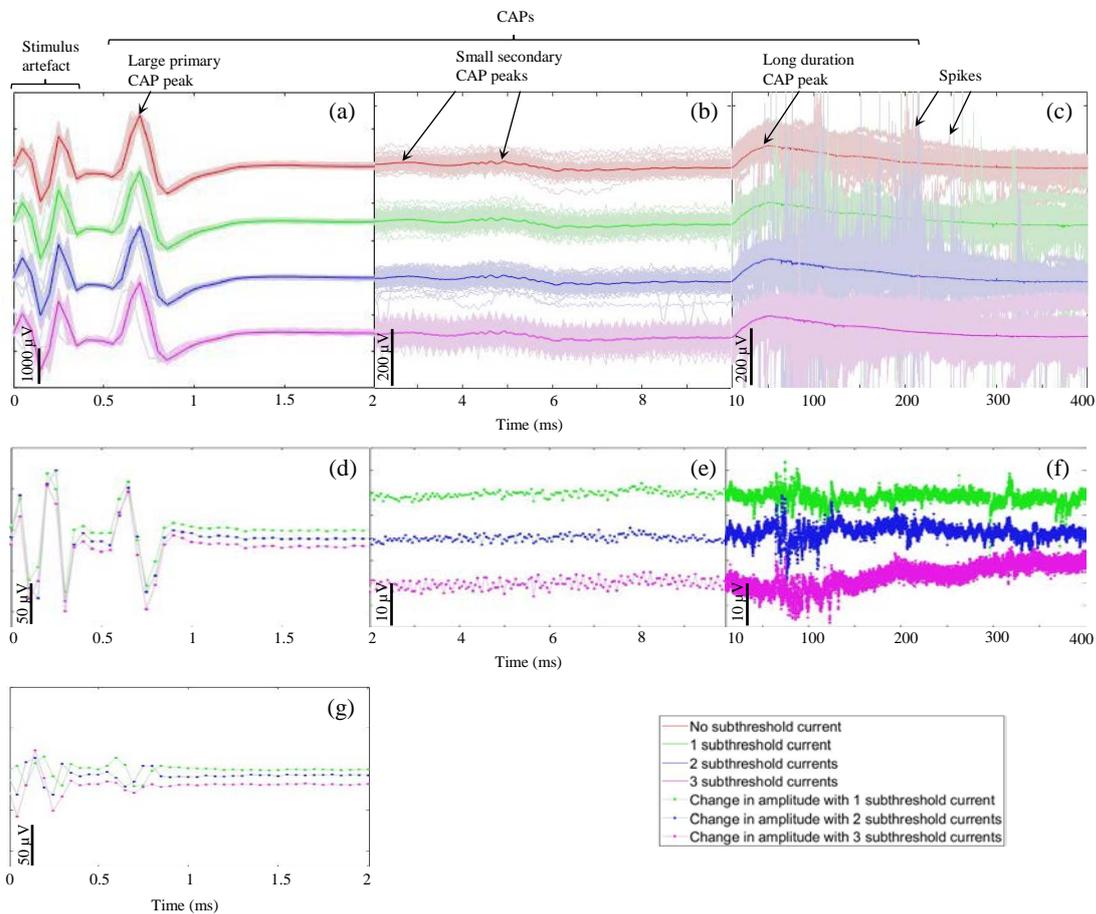

Figure 5: CAPs recorded from the sciatic nerve in response to stimulus applied approximately 70 mm distally on the peroneal nerve with none (red), one (green), two (blue), and three (purple) subthreshold currents applied (a – c). The individual (shaded lines) and mean recordings (bold lines) show a large stimulus artefact between 0 and 0.4 ms, followed a large primary CAP in the 0 to 2 ms window (a), two small secondary CAPs in the 2 to 10 ms window (b) and a long duration CAP coinciding with multiple, short duration spikes in the 10 to 400 ms window (c). Significant differences between the mean recordings are visible for the primary CAP indicating a temporal shift of the CAP in the presence of the subthreshold currents (d), whereas any differences in the secondary and long duration CAPs are below the noise threshold (e-f). The differences between the mean recordings in the primary CAP are significantly reduced with the subthreshold current is reduced from +/- 70 to +/- 30 µA.

When the amplitude of the subthreshold currents was reduced to +/- 30 µV, lag values of -1, 1 and 0 were calculated from cross correlation of the upsampled waveforms within the temporal window of 0.5 to 2 ms, corresponding to a temporal shifts of -0.5, +0.5 and 0 µs in the negative direction. These temporal shifts were again visible in the difference in mean recorded waveforms, although with markedly smaller amplitudes than previously, Fig. 5e.

## 4 Discussion

The first hypothesis, that the subthreshold currents contribute to excitation of the underlying nerve fibres, was confirmed through two observations: (i) that two subthreshold currents applied to the same section of nerve produce sustained extension of the hindshank; and, (ii) that the CAP amplitude increases with application of one subthreshold current, and then increases further with two subthreshold currents, when applied to the same section of nerve. This result is expected given the summative effect of currents on neural membrane excitability, and agrees in principle with stochastic resonance experiments [13, 14], as well as the ephaptic interaction study on rat *in vivo* [28] where the subthreshold component was instead produced by a CAP.

The second hypothesis, that increased excitability would increase synchrony in the neural activity and augment the propagation velocities, was confirmed with low confidence by two observations from the large primary CAP produced in subject 2: (i) a lag of -2.5 to -3 µs in the presence of subthreshold currents; and, (ii) a biphasic difference in the mean waveforms of around 140 µV peak-peak amplitude in the presence of subthreshold currents. While confidence in the lag values is limited by the use of upsampling with spline interpolation and because they represent an increase in propagation velocity of only 0.43 %, they do agree with the statistically significant biphasic difference in the mean waveforms characteristic of a negative temporal shift.

It is not clear whether the observed change in CAP velocity agrees with modelling predictions in [24-27], that ephaptic interactions slow down the CAP, because the models did not consider external current. In the present study, it is conceivable that the partially excited membranes took less time to depolarize, which would produce the observed increase in CAP velocity. If this was the case, however, it is not clear why the lag values did not increase linearly with the number of subthreshold currents applied. Finite element modelling of nerve fibres may provide insight into the observed change in CAP velocity, as well as the best way to configure multiple subthreshold currents to augment such an effect.

Limitations of this study are the low number of subjects (n=2), low temporal sampling rate (20 kHz), high noise, and small range of stimulus and subthreshold current amplitudes and frequencies investigated.

## 5 Conclusion

In this study, we confirmed that administering subthreshold currents increases excitability of the underlying fibres, and, with low confidence, that this increased excitability changes the propagation velocity of neural activity in the underlying fibres. While more work needs to be done in this area to improve confidence in the results, these initial findings contribute to understanding of possible mechanisms of neuromodulation using subthreshold currents.

**Disclosures**

The authors have no relevant financial interests in this article and no potential conflicts of interest to disclose.

**Acknowledgements**

The authors would like to thank David Holder, Kirill Aristovich and Enrico Ravagli, from the EIT Research Group at University College London, for fabrication of electrode arrays used in this study; Darren Svirskis from the School of Pharmacy, The University of Auckland for assistance with PEDOT coating electrodes; and staff at the Faculty of Medical and Health Sciences, The University of Auckland for their help with experiments.